\documentclass[fleqn,usenatbib]{mnras}
\usepackage[utf8]{inputenc}
\usepackage[T1]{fontenc}
\usepackage{graphicx}
\usepackage{amsmath}
\usepackage{amssymb}

\title[Pulsar death line revisited -- II]
{Pulsar death line revisited -- II. 'The death valley'}
\author[V. S. Beskin, A. Yu. Istomin]
{V. S. Beskin$^{1,2}$\thanks{E-mail:
beskin@lpi.ru} and A. Yu. Istomin$^{2}$ \\
$^{1}$P.N.Lebedev Physical Institute, Leninsky prosp., 53, Moscow, 119991, Russia\\
$^{2}$Moscow Institute of Physics and Technology, Dolgoprudny,
Moscow region, Institutsky per. 9, 141700, Russia}

\begin{document}

\date{Accepted, Received}

\pagerange{\pageref{firstpage}--\pageref{lastpage}} \pubyear{2008}

\maketitle

\label{firstpage}

\begin{abstract}
In this paper, which is the second in a series of papers, we analyse what
parameters can determine the width of the radio pulsar ‘death valley’ in the
$P$--${\dot P}$ diagram. Using exact expression for the maximum potential 
drop, which can be realised over magnetic polar caps and the corresponding 
threshold for the secondary plasma production determined in Paper I, we 
analyse in detail the observed distribution of pulsars taking into account 
all the possible parameters (radius $R$  and moment of inertia of a neutron 
star $I_{\rm r}$, high-energy tail in the $\gamma$-quanta energy distribution 
giving rise to secondary particles, etc.) which could broaden 'the death line'. 
We show that the consistent allowance for all these effects leads to a 
sufficiently wide of 'the death valley' containing all the observed pulsars
even for dipole magnetic field of a neutron star.
\end{abstract}


\section{Introduction}
\label{Sect1}

Pulsar radio emission is believed to be produced by a secondary electron-positron 
plasma generated in the polar regions of a neutron star~\citep{sturrock, RS, 
Arons1982, L&GS, L&K}. For this reason, the cessation condition for the generation 
of secondary particles is associated with the so-called 'death line' on the 
$P$--${\dot P}$ diagram, where $P$ is the pulsar period, and ${\dot P}$ is its time
derivative. However, despite in-depth research on the generation of 
secondary plasma conducted since the beginning of the eighties 
of the last century~\citep{dauharding82, GI85, AE2002, Denis, Tim2010, ML2010, 
TimArons2013, PhSC15, TimHar2015, CPhS16} up to now, a large number of different 
options have been discussed in the literature~\citep{RS, blandford76, Arons1982, MelU}, 
leading to markedly different conditions which set 'the death line' of radio
pulsars~\citep{CR93, ZhHM2000, Hib&A, Faucher, KonarDeka}. 

We immediately note that in this series of works, we discuss the 'classical' mechanism 
of particle production only. As is well-known, this process includes primary particle 
acceleration by a longitudinal electric field, $\gamma$-quanta emission due to 
curvature radiation, production of secondary electron-positron pairs, and, finally, 
secondary particles acceleration in the opposite direction, which also leads to the 
creation of secondary particles~\citep{sturrock, RS}. Thus, we do not consider particle 
production due to Inverse Compton Scattering, which, as is well known~\citep{blandford76, 
ZhHM2000, Barsukov07}, can also be a source of hard $\gamma$-quanta. As an excuse, we 
note that first of all, we will be interested in old pulsars, in which the surface
temperature may not be high enough to form a sufficient number of X-ray photons. 

Moreover, we also do not include into consideration synchrotron photons emitted by 
secondary pairs. The point is that the energy of synchrotron photons emitted by 
secondary particles is approximately $15$--$20$ times less than the energy of 
curvature photons emitted by primary particles (see, e.g.~\citealt{GI85, Denis}). 
Therefore, near the threshold for particle production, when the free path lengths 
of curvature photons becomes close to the radius of the star $R$, the pulsar 
magnetosphere turns out to be transparent for synchrotron photons.

As a result, as was first shown by~\citet{RS}, the cessation condition for the 
pair creation determining the position of 'the death line' on the $P$--${\dot P}$
diagram can be evaluated from the equality of the height of the 1D vacuum gap 
\begin{equation}
     H_{\rm RS} \sim 
     \left(\frac{\hbar}{m_{\rm e}c}\right)^{2/7}
     \left(\frac{B_{0}}{B_{\rm cr}}\right)^{-4/7}
     R_{\rm L}^{3/7} R_{\rm c}^{2/7}
\label{HRS}    
\end{equation}
and the polar cap radius
\begin{equation}
    R_{\rm cap} \approx \left(\frac{\Omega R}{c}\right)^{1/2} R. 
\label{RRS}    
\end{equation}
Here, $B_{\rm cr} = m_{\rm e}^2c^3/e\hbar = 4.4 \times 10^{13}$ G is
the Schwinger magnetic field, $R_{\rm L} = c/\Omega$ is the radius
of the light cylinder ($\Omega = 2\pi/P$ is the star angular velocity), 
and $R_{\rm c}$ is the curvature of magnetic field lines near the
magnetic pole. For magneto-dipole energy losses
\begin{equation}
    W_{\rm tot} \sim \frac{B_{0}^{2}\Omega^{4}R^{6}}{c^3}
\label{Wtot0}    
\end{equation}
and the dipole magnetic field stricture, when
\begin{equation}
   R_{\rm c} = \frac{4}{3} \, \frac{r}{\theta_{m}},
\label{Rc11}    
\end{equation}
($r$ and $\theta_{m}$ are the polar coordinates relative to the magnetic axis, 
$r$ is the distance from the star centre), one can obtain for 'the death line'~\citep{RS}
\begin{equation}
  {\dot P}_{-15} = \beta_{\rm d} P^{11/4},
\label{dl1}  
\end{equation}
where ${\dot P}_{-15} = 10^{15}{\dot P}$ and $\beta_{\rm b}^{\rm RS} \sim 1$.

\begin{figure}
		\center{\includegraphics[width=0.9\linewidth]{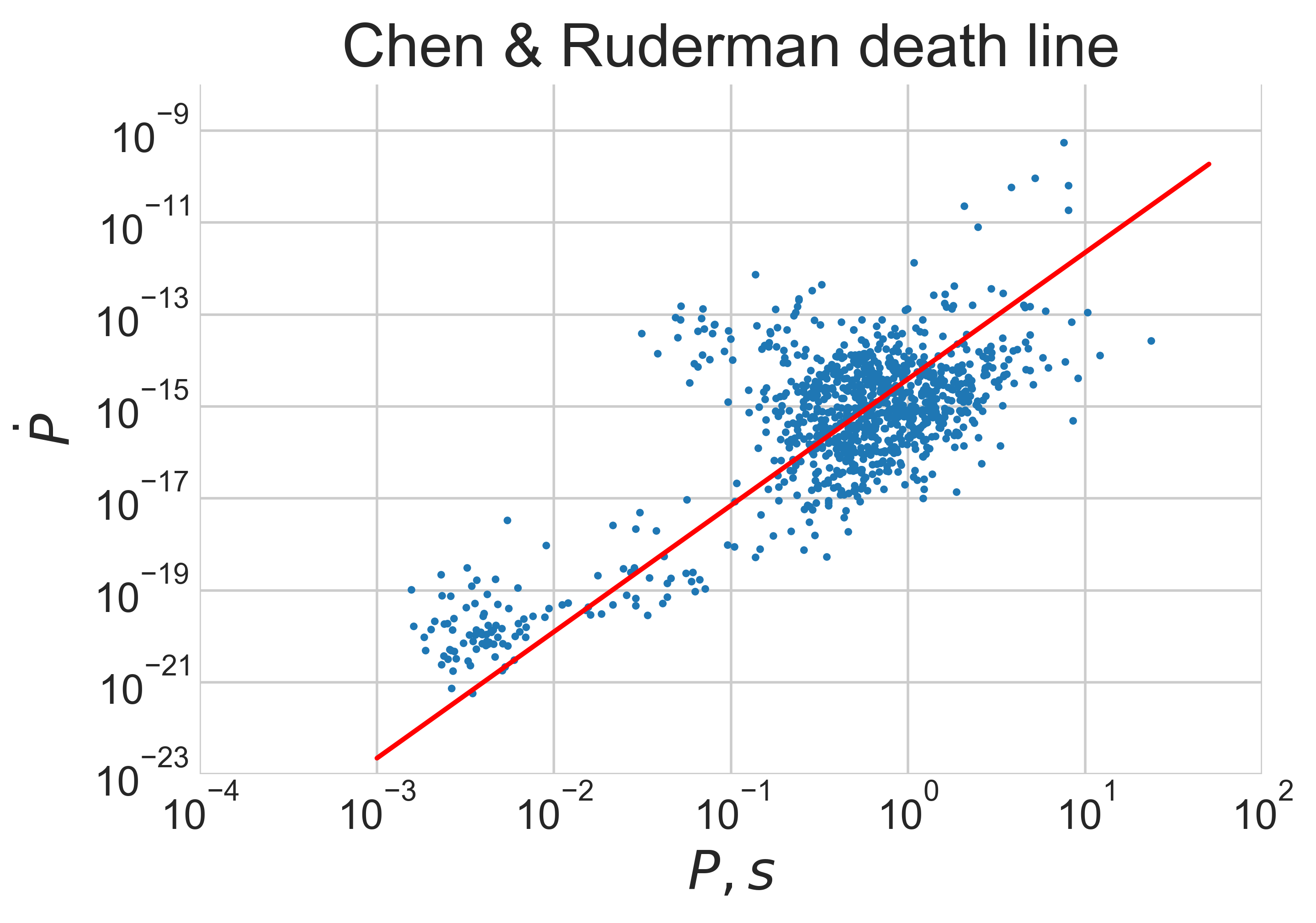}}
	\caption{$P$--${\dot P}$ diagram taken from the ATNF catalogue~\citep{ATNF}.
	The line corresponds to relation (\ref{dl1}) with \mbox{$\beta_{\rm d} = 4$}
	obtained by~\citet{CR93} for dipole magnetic field.}
\label{fig1}	
\end{figure}

It is clear that in the mid-70s such accuracy was quite acceptable,
especially since expression (\ref{dl1}) really limited from below
most of the pulsars in the $P$--${\dot P}$ diagram. However, as was 
already noted, at present this issue requires substantial revision. 
Indeed, as one can see form Figure~\ref{fig1}, there are many radio pulsars 
below 'the death line' drawn for characteristic values, i.e. for 
neutron star radius $R =10$ km and magnetic field $B_{0}= 10^{12}$ G 
($\beta_{\rm d} = 4$ according to~\citealt{CR93}).
As can be seen from Table~\ref{tab0}, the derivative of the period for some 
pulsars turns out to be 1-2 orders of magnitude less than that of
the classical Ruderman-Sutherland death line, i.e. gives: 
$\beta_{\rm d} = (0.01$--$0.1)\beta_{\rm d}^{\rm RS}$.

\begin{table*}
\caption{Pulsars deep below 'the death line' ($\beta_{\rm d} < 0.02$) taken from the ATNF catalogue~\citep{ATNF}. 
See the text for more detail.}
\begin{tabular}{cccccccccc}
\hline
PSR & $P$  & ${\dot P}_{-15}$ & $B_{12}^{\rm ATNF}$ &$B_{12}^{\rm MHD}$ & $B_{12}^{\rm BGI}$ & $\Lambda$ & ${\cal K}$ & $\xi$    & $\beta_{\rm d}$ \\
    & (s)  &    &    &  $(\chi = 60^{\circ})$ &  $(\chi = 60^{\circ})$ &  &   &  &  \\
\hline
%J0156$+$3949   &  1.81  &  0.15  & &&&&&&& $-$ & $-$  &  0.095 \\
J0250$+$5854    &  23.53 &  27.16 & 25.66 & 25.88 & 42.80 & 39 & $1.5\times 10^5$ & 7.1  &  0.003 \\
J0343$-$3000    &  2.60  &  0.06  & 0.39 & 0.39 & 0.65 & 37 & $1.1\times 10^6$ & 8.6 &  0.004 \\
J0418$+$5732    &  9.01  &  4.10  & 6.17 & 6.22 & 10.29 & 35 &   $7.3\times 10^4$& 6.6  &  0.010 \\
J0457$-$6337    &  2.50  &  0.21  & 0.74 & 0.74 & 1.23 & 33 & $1.3\times 10^5$ & 7.1  &  0.017 \\
J0656$-$2228    &  1.23  &  0.03  & 0.18 & 0.19&  0.31 & 33 & $3.5\times 10^5$  & 7.8 &  0.015 \\
 J0901$-$4046    &  75.89  &   215.  &  128 &  129 &  214 &  42 &  $1.2\times 10^5$ &  7.9  & 0.001 \\
J0919$-$6040    &  1.22  &  0.01  & 0.11 & 0.11 & 0.19 & 36 & $1.5\times 10^6$ & 8.9  &  0.006 \\
J1210$-$6550    &  4.24  &  0.43  & 1.37 & 1.38 & 2.29 & 35 & $ 2.2\times 10^5$ & 7.3 &  0.008 \\
J1232$-$4742    &  1.87  &  0.01  & 0.16 & 0.17& 0.27 & 39 & $3.3\times 10^6$ & 9.4 &  0.003 \\
J1320$-$3512    &  0.46  &  0.002  & 0.03 & 0.03 & 0.05 & 32 & $9.0\times 10^5$ & 8.4 &  0.017 \\
J1333$-$4449    &  0.46  &  0.0005  & 0.01 & 0.01 & 0.02 & 34 &  $3.2\times 10^6$ & 9.4  &  0.009 \\
J1503$+$2111    &  3.32  &  0.14    & 0.69 & 0.71 & 1.15 & 37 & $5.8\times 10^5$ & 8.1  &  0.005 \\
J1638$-$4344    &  1.12  &  0.02    & 0.17& 0.17 & 0.28 & 32 & $2.9\times 10^5$ & 7.6 &  0.018 \\
%J1700$-$3919    &  0.56  &  0.005  &&&&&&&14 & 6.7  &  0.025  \\ 
%J1710$-$2616    &  0.95  &  0.02   &&&&&&& 87(?) & 32(?)  &  0.023  \\ 
%J1717$-$3953    &  1.08  &  0.03   &&&&&&& 299(?) & 104(?) &  0.024  \\
J1801$-$1855    &  2.55  &  0.18    & 0.69 & 0.69 & 1.15 & 33 & $1.8\times 10^5$ & 7.3   &  0.014  \\
J1805$-$2447    &  0.66  &  0.006   & 0.06 & 0.06 & 0.11& 32 & $5.1\times 10^5$ & 8.0    &  0.019  \\
%J1806$-$1920    &  0.88  &  0.017  &&&&&& &660(?) & 253(?) &  0.024  \\
%J1834$-$1202    &  0.61  &  0.007   &&&&&&& 63  & 29     &  0.026  \\
J1859$+$7654    &  1.39  &  0.05    & 0.27 & 0.27 & 0.45 & 32 & $2.0\times 10^5$ & 7.4    &  0.020  \\
%J1901$+$0621    &  1.39  &  0.007  &&&&&&&  43  & 17    &  0.030  \\
%J1901$+$1306    &  0.83  &  0.02   &&&&&&&  35  & 9.3   &  0.025  \\
J1915$+$0752    &  2.06  &  0.14    & 0.55 & 0.55 & 0.91 & 32 & $1.4\times 10^5$ & 7.1  &  0.019  \\
J1954$+$2923    &  0.43  &  0.0002   & 0.01 & 0.01 & 0.02 & 39 & $2.3\times 10^7$ & 11.0 &  0.002  \\
J2136$-$1606    &  1.23  &  0.16    & 0.14 & 0.14 & 0.24 &  34 & $7.6\times 10^5$ & 6.6   &  0.009  \\
J2144$-$3933    &  8.51  &  0.50    & 2.09 & 2.10 & 3.48 &  41 & $1.5\times 10^6$ & 8.9   &  0.001  \\
J2251$-$3711    &  12.12  &  13.10  & 12.74 & 12.85 & 21.25& 34 & $3.1\times 10^4$ & 6.0  &  0.014  \\
J2310$+$6706    &  1.94  &  0.08    & 0.39 & 0.39 & 0.65 &  34 & $2.9\times 10^5$ & 7.4   &  0.012  \\  
\hline
\end{tabular}
\label{tab0}
\end{table*}

On the other hand, it is important that their number decreases with the 
distance from it. In total, there are 110 pulsars with $\beta_{\rm d} < 0.1$, 
and only 21 pulsar with $\beta_{\rm d} < 0.02$. This implies that in reality
we deal with 'the death valley' corresponding to the tail of the 
distribution with respect to some parameters. Therefore, one of the main 
tasks of our consideration is the question of which parameter leads to 
a decrease in the observed deceleration rate ${\dot P}$.

The idea of 'the death valley' is not new. It was first discussed by~\citet{CR93}, 
who introduced this term, and then this issue was discussed in many other works 
(see, e.g.~\citealt{ZhHM2000, Gonthier, KouTong}). In particular, the authors 
discussed a possible role of a nondipole magnetic field. However, none of these 
works, based on qualitative estimates, studied quantitatively the question of 
the real 3D structure of the particle acceleration region, not to say about the
spread of such parameters as the masses and moments of inertia of neutron stars. 

To clarify this issue, in Paper I~\citep{PaperI}, we set ourselves a task to reconsider 
all the basic approximations which are usually used in constructing of the model the 
secondary plasma generation, but which may work poorly near 'the death line'. These 
refinements concerned the electric potential, the influence of the emission spectrum 
of primary particles, and the effects of general relativity. Such a detailed
study has never been done before.

As a result, the conditions for the cascade generation of particles were formulated, 
which we will consider here as a condition which determines 'the death line' on the 
$P{\dot P}$-diagram. Let us emphasize that as both relativistic corrections and the 
connection between the deceleration rate ${\dot P}$ and the magnetic field depend
on the radius $R$ and moment of inertia $I_{\rm r}$, we, in fact, deal with a
rather wide 'death valley', i.e. with a sufficiently wide area whose width depends 
on the spread of these values. Determining the real width of 'the death valley', 
as well as explaining the existence of radio pulsars with extremely low deceleration 
rates is the main goal of this work.

 We emphasize once again that the main goal of this work was to demonstrate that 
the original Ruderman-Sutherland idea of the death line (dipole magnetic field, vacuum 
gap) leading to dependence (\ref{dl1}) is in good agreement with observations. In other
words, we show below that the agreement is achieved even within the framework of the 
power dependence ${\dot P} \propto P^{11/4}$, since the simultaneous taking into account 
all the effects mentioned above reduces significantly the coefficient $\beta_{\rm b}$. 
Thus, comparison with other models is beyond the scope of this article.

For this reason, in this paper we consider only a dipole magnetic field, despite a 
large number of works which indicated that it is impossible to explain 'the death 
line' in a dipole magnetic field~\citep{Arons93, AKh2002, BTs2010, IEP2016, Bilous}. 
 In particular, we do not discuss the model with a fixed curvature radius
$R_{\rm c} = R$, also considered by~\citet{CR93}. By the way, taking into account 
the effects discussed in Paper I  this boundary (it corresponds to dependence 
${\dot P} \propto P^2$) should be located well below the observed pulsars.

As an additional argument, we can cite a sufficiently large number of pulsars with 
drifting subpulses~\citep{drift1,drift2}, for which, in the framework of the carousel 
model~\citep{carousel1, carousel2, carousel3}, a regular axisymmetric magnetic field 
is required. Moreover, it is precisely in the region of plasma generation, since it 
is this region that determines the drift velocity. Such a configuration is hardly 
possible for a random orientation of the nondipole component. Of course, individual 
pulsars can have a significant non-dipole magnetic field (for example, as a pulsar 
PSR J0030+0451, see~\citealt{NICER} for more detail).

%In other words, one of our tasks is to verify the possibility of explaining the position 
%of 'the death line' by other factors which are usually not taken into account when analysing 
%the processes of production of secondary plasma.

The paper is organized as follows. In Section 2, we present a summary of the main results obtained 
in Paper I. They refer to all possible amendments which have not yet been taken into account together. 
In addition, the parameters of two evolutionary scenarios are formulated in what follows. Further, 
in Section 3, the real boundaries of ''the death valley''  are determined, which are in good agreement 
with the observations. Then, after discussing the nature of the knee in 'the death line' in Section 4,
a discussion of the results is given in Section 5.

\section{Basic equations}

\subsection{Paper I --- general results}

At first, in Paper I, we assumed that due to time irregularity of the secondary plasma 
production~\citep{Tim2010,TimArons2013, TimHar2015, PhTS20}, 
almost the entire region of open field lines can be considered in a vacuum approximation:
$\rho_{\rm e} = 0$. Using this approximation, we constructed an exact three-dimensional 
solution for longitudinal electric field $E_{\parallel}$ in the polar regions of a 
neutron star
\begin{eqnarray}
&&E_{\parallel} = -\frac{1}{2} \, \frac{\Omega B_{0}R_{0}}{c}\cos\chi \times
\nonumber \\
&&\sum_{i} c_{i}^{(0)} \lambda_{i}^{(0)}\left(\frac{r}{R}\right)^{-\lambda_{i}^{(0)}/\theta_{0}-1}
J_{0}(\lambda_{i}^{(0)} \theta/\theta_{0})  
\label{Epar1}   \\
&&    - \frac{1}{4} \, \frac{\Omega B_{0}R_{0}}{c}\, \frac{R_{0}}{R}
\sin\varphi \sin\chi \times
\nonumber \\
&&\sum_{i} c_{i}^{(1)} \lambda_{i}^{(1)} \left(\frac{r}{R}\right)^{-\lambda_{i}^{(1)}/\theta_{0}-1}
J_{1}(\lambda_{i}^{(1)} \theta/\theta_{0})
\nonumber \\
&&-\frac{3}{16} \, \left(\frac{f}{f_{\ast}}\right)^{1/2} 
    \left(1 - \frac{f}{f_{\ast}}\right) \, \frac{\Omega B_{0}R_{0}^3}{c R^2} 
    \left(\frac{l}{R}\right)^{-1/2}  \sin\varphi \sin\chi.
\nonumber    
\end{eqnarray}
Here, $R$ is the star radius, $B_{0}$ is the magnetic field at the star magnetic pole,
\begin{equation}
R_{0} = f_{\ast}^{1/2} \left(\frac{\Omega R}{c}\right)^{1/2} 
\end{equation} 
is the polar cap radius, $f_{\ast} \approx 1$ is the standard dimensionless 
polar cap area, and $l$ is the distance along the magnetic field line 
$f =$ const. Finally, $\lambda_{i}^{(0,1)}$ are the zeros of the Bessel 
functions $J_{0,1}(x)$, and the expansion coefficients $c_{i}^{(0,1)}$ 
satisfy the conditions
\begin{equation}
    \sum c_{i}^{(0)}J_{0}(\lambda_{i}^{(0)} x) = 1 - x^2   
\label{sum1}
\end{equation}
\begin{equation}
    \sum c_{i}^{(1)}J_{1}(\lambda_{i}^{(1)} x) = x - x^3.   
\label{sum2}
\end{equation}
Accordingly, the potential drop $\psi(r_{m}, \varphi_{m})$ over the polar cap with 
the polar coordinates $r_{m}, \varphi_{m}$ on the scale $l \sim R_{0}$  can be written 
down as  
\begin{eqnarray}
\psi(r_{m}, \varphi_{m})  = \frac{1}{2} \, 
\frac{\Omega B_{0}R_{0}^2}{c}\left(1-\frac{r_{m}^2}{R_{0}^2}\right)\cos\chi 
\label{psis} \\
 + \, \frac{3}{8} \, \frac{\Omega B_{0}R_{0}^2}{c}\, 
\frac{r_{m}}{R}\left(1-\frac{r_{m}^2}{R_{0}^2}\right)\sin\varphi_{m} \sin\chi.
\nonumber
\end{eqnarray}
Knowing now the longitudinal electric field $E_{\parallel}$ (\ref{Epar1}), we 
can determine the production rate of secondary particles at sufficiently large 
periods $P$. 

Note that as one can see from (\ref{Epar1}), in real dipole geometry, 
for non-zero inclination angles $\chi$, the longitudinal electric field  does not 
vanish on the scale $l \sim R_{0}$, which was previously assumed by~\citet{MTs92}. 
It decreases much more slowly, as $\propto (l/R)^{-1/2}$. This effect, however, 
is significant only for almost orthogonal rotators due to the additional factor
$R_{0}/R$. 

Next, the corrections related to the effects of general relativity were 
taken into account. First of all, as is well known~\citep{B90, MTs92, PhSC15, PhTS20}, 
the effects of general relativity increase the electric potential (and, hence, the 
particle energy) as
\mbox{$\psi_{GR} =  K_{\psi}\psi$}, where 
\begin{equation}
 K_{\psi} = \left(1 - \frac{\omega}{\Omega}\right)\left(1 - \frac{r_{\rm g}}{R}\right)^{-1}.
\label{GR11}
\end{equation}
Here, $r_{\rm g} = 2GM/c^2$ is the gravitational radius, and 
\begin{equation}
\frac{\omega}{\Omega} = \frac{I_{\rm r}r_{\rm g}}{M R^3},
\label{GR5}
\end{equation}
where $\omega$ is the Lense-Thirring angular velocity ($M$ and $I_{\rm r}$ are the neutron 
star mass and moment of inertia, respectively). However, to determine all the 
characteristics of particle production, we also need the corrections to the curvature 
radius of the magnetic field line $R_{\rm c}$ as well as to the polar cap radius
$R_{0}$: $R_{{\rm c, GR}} = K_{\rm cur}R_{\rm c}$ and $R_{0, {\rm GR}} = K_{\rm cap}R_{0}$. 
As was shown in Paper I, they look like
\begin{equation}
K_{\rm cur} = 1 - \frac{1}{2}\frac{r_{\rm g}}{R},
\label{Rcur}
\end{equation}
\begin{equation}
K_{\rm cap} = 1 - \frac{3}{8}\frac{r_{\rm g}}{R}.
\label{Kcap}
\end{equation}
Finally, the magnetic field on the star surface $B_{0}$, due to a well-known correction to the
magnetic flux~\citep{Ginzburg}, increases as $B_{0,{\rm GR}} = K_{B}B_{0}$, where
\begin{equation}
K_{B} = 1 + \frac{3}{4}\frac{r_{\rm g}}{R}.
\label{KB}
\end{equation}
Note that such an increase in the magnetic field takes place if we fix its asymptotic behaviour 
at large distances from the neutron star. As will be shown below, it is precisely this case that 
is of interest.

%Further, it was shown that, close to ''the death line'', a major role begin to play those
%$\gamma$-quanta whose energy is several times greater than that commonly used characteristic 
%energy of curvature radiation $\hbar \omega_{\rm c}$, where
Further, it was shown that the secondary particles generated at the smallest distance 
from the place of $\gamma$-quanta radiation, correspond to the $\gamma$-quantum energy, which 
significantly exceeds the characteristic energy of the curvature radiation $\hbar \omega_{\rm c}$, 
where
\begin{equation}
\omega_{\rm c} = \frac{3}{2} \, \frac{c}{R_{\rm c}} \, \gamma_{\rm e}^3.
\label{KBomega}
\end{equation}
Denoting this energy as $\xi \hbar \omega_{\rm c}$, it was shown that the values of 
$\xi$ are to be determined from the relation
\begin{equation}
\xi^{5/2} \, e^{\xi} \left(1 - \frac{55}{72} \, \frac{1}{\xi} + \dots \right) = {\cal K},
    \label{49}
\end{equation}
where 
\begin{equation}
{\cal K} = \frac{4\sqrt{2}}{3\sqrt{3 \pi} \Lambda} \, \frac{B_{\rm cr}}{B}  
 \, \frac{R_{\rm c}}{a_{\rm B}}  \, \gamma_{\rm e}^{-2} \approx
 40 \, R_{\rm c, 7} B_{12}^{-1}\gamma_{7}^{-2}. 
    \label{50}
\end{equation}
Here, $B_{\rm cr} = m_{\rm e}^2c^3/e\hbar \approx 4.4 \times 10^{13}$  G is the 
critical magnetic field, $a_{\rm B} = \hbar^2/m_{\rm e}e^2 = 5.3 \times 10^{-9}$ cm 
is the Bohr radius, and $\Lambda = 15$--$20$ is the logarithmic factor: 
$\Lambda \approx \Lambda_{0} - 3\ln\Lambda_{0}$, where
\begin{equation}
\Lambda_{0} = \ln\left[
\frac{e^2}{\hbar c}\,\frac{\omega_B R_{\rm c}}{c}
\left(\frac{B_{\rm cr}}{B}\right)^2
\left(\frac{m_{\rm e}c^2}{{\cal E}_{\rm ph}}\right)^2\right].
   \label{48}
\end{equation}
Accordingly, $R_{\rm c, 7} = R_{\rm c}/(10^7 {\rm cm})$, where $R_{\rm c}$ 
is the curvature radius of the magnetic field lines, and $\gamma_{7} = \gamma_{\rm e}/10^7$. 

The corresponding values of $\xi$ are also given in Table~\ref{tab0}. 
In this case, the Lorentz-factor of primary particles was determined
as $\gamma_{\rm e} = e\psi/m_{\rm e}c^2$, where $\psi$ was taken by 
the relation (\ref{psis})  for $r_{m} = 0.7 \, R_{0}$. Accordingly,
the curvature radius \begin{equation}
R_{\rm c} = \frac{4}{3} \, \frac{R^2}{r_{m}}
   \label{Rc}
\end{equation}
was taken for the same distance $r_{m}$. As we see, the values of $\xi$ 
for the pulsars near 'the death line' turned out to be large enough. Thus, 
taking this correction into account is also important for the pulsars 
located in 'the death valley' region.

Let us finally formulate the condition for the existence of the cascade production of 
particles, which we will consider as the condition which determines the position of 
'the death valley'. First of all, note that the beginning of the cascade (and, hence, 
the filling of this region with a secondary electron-positron plasma) can be initiated
by the cosmic gamma background, which, as is known, leads to $10^5$--$10^8$ primary 
particles per second in the polar cap region~\citep{SRad82}. It is clear that for the
cascade production of secondary plasma in the open magnetic field lines region, it is
necessary not only to produce particles by $\gamma$-quanta propagating from the pulsar
surface (this process can take place up to heights of $H \sim R$, i.e. on the scale of the 
diminishing of the dipole magnetic field). It is necessary that the secondary particles 
return to the region of a strong longitudinal electric field, accelerate, emit hard 
$\gamma$-quanta, which would have time to give birth to secondary particles above the 
surface of the neutron star. 

As for the return of secondary particles to the pulsar surface from the region $H \sim R$, 
then, as noted previously, it can be easily explained by the slowly decreasing longitudinal 
field mentioned above. On the other hand, a particle moving toward the neutron star surface 
will be able to acquire the required energy only at a height of $H \sim R_{\rm cap} \sim 0.01 R$. 
Accordingly, the free path length of a $\gamma$- quantum should be of the same order. Therefore, 
it is the condition for the production of secondary particles above the very surface of the pulsar 
that should be considered as the condition for the existence of a cascade.

According to the results in Paper I, the condition for the existence 
of a cascade can be written as $P < P_{\rm max}$, where
\begin{equation}
P_{\rm max} = 0.7 \, \xi^{2/15} \frac{K_{\psi}^{2/5}}{K_{\rm cur}^{4/15}}
f_{1.6}^{3/5}
\Lambda_{15}^{2/15}
R_{12}^{19/15}
B_{12}^{8/15}x_{0}^{4/15} {\cal P}^{2/5} \, {\rm s}.
\label{Pmax}
\end{equation}
Here, $f_{1.6} = f_{\ast}/1.6$, $\Lambda_{15} = \Lambda/15$,
$R_{12} = R/(12 \, {\rm km})$, and $I_{100} = I_{\rm r}/(100 \, M_{\odot} {\rm km}^2)$.
The choice of such a normalization for the moment of inertia $I_{\rm r}$ is due to 
the fact that we will further use the results obtained by~\citet{Greif20}, in which 
$I_{\rm r}$ is presented just in this form.
Finally, the last two parameters in (\ref{Pmax}), $x_{0} = r_{m}/R_{0}$  and
\begin{equation}
{\cal P}(r_{m}, \varphi_{m}) = \left(\cos\chi
+ \frac{3}{4}x_{0}\frac{R_{0}}{R}\sin\chi\cos\varphi_{m}\right) (1 - x_{0}^2),
\label{b3}
\end{equation}
determine the dependence of the ignition condition on the position on the polar cap. 
Unlike in Paper I, here we explicitly write down the dependencies on all possible
parameters.

\subsection{Two evolutionary scenarios}

It is clear that expression (\ref{Pmax}) is still not enough to define 
'the death line' in the $P$--${\dot P}$ diagram. For doing this, we need to 
express the magnetic field $B_{0}$ in terms of the observed quantities. In 
other words, we need to specify a braking model of radio pulsars.

Below we consider two braking models. According to the most popular model based 
on the results of numerical simulations~\citep{spitkovsky, cont2012, TchPhS16}, we have
\begin{equation}
\dot P_{\rm MHD}  =  \frac{\pi^2}{P} \frac{B_{0}^{2}R^{6}}{I_{\rm r} c^{3}}(1+\sin^{2}\chi).
\label{MHD_P}
\end{equation}
On the other hand, according to the semi-analytical model proposed by~\citet{BGI93},
for the pulsars near 'the death line', we can write down
\begin{equation}
\dot P_{\rm BGI} = \frac{\pi^2 f_{\ast}^2}{P} \frac{B_{0}^{2}R^{6}}{I_{\rm r} c^{3}}\left(\cos^2\chi + {\cal C}\right).
\label{BGI_P}
\end{equation}
Here,
\begin{equation}
{\cal C} = k\left(\frac{R_{0}}{R}\right)^{1/2} = \varepsilon P^{-1/2}
\label{CalC}
\end{equation}
($P$ is in seconds), $k \sim 1$, and $\varepsilon$ belongs to the range between 0.005 and 0.02~\citep{novoselov}. 
However, the last term in (\ref{BGI_P}) plays a role only for orthogonal pulsars, 
which we do not consider here.

The corresponding magnetic fields, determined by relations (\ref{MHD_P})--(\ref{BGI_P}), are also shown
in Table~\ref{tab0} for the characteristic values $R = 12$ km, $I_{\rm r} = 100 \, M_{\odot}$km$^{2}$ and
$\chi = 60^{\circ}$. As one can see, for these parameters, the magnetic fields $B^{\rm MHD}$ practically 
coincide with the values given in the ATNF catalogue~\citep{ATNF}. On the other hand, the magnetic fields 
for the BGI model turn out to be twice as large.

Note that since the energy losses $J_{\rm r} \Omega {\dot \Omega}$ (and, therefore, 
the measured value of ${\dot P}$) depend on a magnetic field at large distances from 
a pulsar, the magnetic field $B_{0}$ on the neutron star surface should indeed
be corrected according to relation (\ref{KB}). As a result, due to the same
dependence of ${\dot P}$ on $P$ and $B_{0}$, we again obtain in both cases 
\mbox{${\dot  P}_{-15} = \beta_{\rm d} P^{11/4}$} (\ref{dl1}), where now
\begin{eqnarray}
\beta_{\rm d}^{\rm MHD} = 2.1 \,\xi^{-1/2} K_{\rm GR} 
f_{1.6}^{-9/4}
\Lambda_{15}^{-1/2}
R_{12}^{5/4}
I_{100}^{-1}h(x_{0})
F^{\rm MHD},
\label{betaMHD} \\
\beta_{\rm d}^{\rm BGI} =  0.8 \,\xi^{-1/2} K_{\rm GR} 
f_{1.6}^{-17/4}
\Lambda_{15}^{-1/2}
R_{12}^{5/4}
I_{100}^{-1}h(x_{0})
F^{\rm BGI}. 
\label{betaBGI}
\end{eqnarray}
Here, the coefficient 
\begin{equation}
K_{\rm GR}  = \frac{K_{\rm cur}}{K_{B}^2 K_{\psi}^{3/2}},
\label{KGR}
\end{equation}
describes the general relativity correction. Since $K_{\rm GR} < 1$, this coefficient, together 
with the parameter $\xi >1$, decreases the value of ${\dot P}$. Finally, the functions $F(x_0, \chi)$, where 
\begin{eqnarray}
F^{\rm MHD}(\chi) = \frac{(1 + \sin^2\chi)}{(\cos\chi + 3/4 \, x_{0} \, (R_{0}/R) \, \sin\chi\cos \varphi_{m})^{3/2}},
\label{FMHD}\\
F^{\rm BGI}(\chi) =  \frac{\cos^2\chi + k(R_{0}/R)}{[\cos\chi + 3/4 \, x_{0} \, (R_{0}/R) \, \sin\chi\cos \varphi_{m}]^{3/2}}
\label{FBGI}
\end{eqnarray}
and
\begin{equation}
h(x_{0}) = x_{0}^{-1}(1 - x_{0}^2)^{-3/2},
\label{fx}
\end{equation}
describe the dependence on the distance from the magnetic axis $x_{0} = r_{m}/R_{0}$ and on the inclination angle $\chi$.

%\begin{figure}
%		\center{\includegraphics[width=0.7\linewidth]{fig2.png}}
%	\caption{Broken ''death line'' obtained for two models of pulsar braking. Dashed lines correspond to ''classical'' death line presented by~\citet{CR93}.}
%\label{fig2}	
%\end{figure}

\section{'The Death Valley'}

%\subsection{Qualitative Consideration}

At the beginning, let us discuss qualitatively whether an accurate allowance for all the possible corrections reduce the value of $\beta_{\rm d}$ enough to explain the entire width of 'the death valley'. First, as we see, numerical coefficients in  expressions (\ref{betaMHD}) and (\ref{betaBGI}) turn out to be less than the initial rough estimate \mbox{$\beta_{\rm d} = 4$} obtained by~\citet{C&R}, especially for the BGI model. This is due to the fact that we used the exact value of the potential drop $\psi$, moreover, in the case when the plasma in the region of the open field lines is completely absent.

Next, according to Table~\ref{tab0}, the photon energy correction $\xi$ reaches values of 7--10, so that for the pulsars located within 'the death valley', the correction factor $\xi^{-1/2}$ turns out to be of the order of 0.3. Further, the general relativistic correction $K_{\rm GR}$ (\ref{KGR}) for the characteristic values ($M = 1.4 \, M_{\odot}$, \mbox{$R = 12$ km,} \mbox{$I_{\rm r} = 100 \, M_{\odot}$ km$^{2}$)} gives $K_{\rm GR} \approx 0.3$. Below, we discuss this issue in more detail, taking into account all the terms, including $R$ and $I_{\rm r}$. But already here one can conclude that the last two factors lower the value of $\beta_{\rm d}$ by an order of magnitude. Thus, this preliminary analysis is enough to conclude that the key parameter $\beta_{\rm d}$ may be significantly less than it is usually assumed.

Thus, our qualitative discussion shows that the consistent inclusion of the above corrections really allows one to significantly shift down 'the death line' in the $P$--${\dot P}$ diagram. Below, we further discuss this issue, trying to understand whether all the pulsars found in 'the death valley' can be explained within the framework of our approach.

Now we proceed to a detailed study of all the quantities included in expressions (\ref{betaMHD})--(\ref{betaBGI}). At first, let us discuss the question of how the parameters of a neutron star, such as their radius $R$, mass $M$ and moment of inertia $I_{\rm r}$, can affect the value of the parameter $\beta_{\rm b}$. At the same time, when analysing the possible scatter in these quantities, we use the results obtained by~\citet{Greif20}, where the corresponding theoretical values are presented.

\begin{table}
\caption{Tabulation of the factor $K_{\rm g}$ (\ref{Kg}).}
\begin{tabular}{cccccc}
\hline
$M \, (M_{\odot})$  & 0.5 & 1.0 & 1.5 & 2.0 & 2.5 \\
\hline
$R = 10$ km & 4.91 & 0.68 & 0.24 & 0.11 & $-$ \\
$R = 11$ km & 4.78 & 0.90 & 0.30 & 0.13 & 0.07 \\
$R = 12$ km & 4.45 & 1.12 & {\bf 0.35} & 0.15 & 0.07 \\
$R = 13$ km & 3.81 & 1.06 & 0.41 & 0.16 & 0.08 \\
$R = 14$ km &  $-$ & $-$  & $-$ & 0.17 & 0.09 \\
\hline
\end{tabular}
\label{tab1}
\end{table}

Table~\ref{tab1} shows the values of factor $K_{\rm g}$
\begin{equation}
K_{\rm g} = K_{\rm GR}R_{12}^{5/4}I_{100}^{-1},
\label{Kg}
\end{equation}
which contains complete information about the role of these parameters. As one can see, for massive neutron stars \mbox{($M \approx 2 \, M_{\odot}$),} the reduction factor can be as small as 0.1 or even smaller. As for the tail of this distribution, the difference between the smallest values of $K_{\rm g}$ and its average value (marked in bold) is only 0.2-0.3. 

Next, we note a strong dependence of $\beta_{\rm b}$ on $f_{\ast}$ for both braking models. As was shown by~\citet{bgi83} and confirmed recently by~\citet{TchPhS16} (see also~\citealt{Gralla})
\begin{equation}
f_{\ast} \approx  f_0 (1 + 0.2 \, \sin^2\chi),
\end{equation}
when $f_0 = 1.4$--$1.6$. Therefore, for angles $\chi$ close to $90^{\circ}$, we have $f_{\ast} = 1.7$--$2.0$. As a result, for the limit value $f_{\ast} = 2$, we get a reduction factor of 0.6 for the MHD model and 0.4 for the BGI model. But also for a more realistic case $f_{\ast} = 1.8$, we have 0.75 for the MHD model and 0.6 for the BGI model. In general, as one can see, the position of 'the death line' depends very much on $f_{\ast}$ (i.e., on the radius of the polar cap $R_{0}$). Below, we discuss this issue in greater detail.

Further, despite the low power $1/2$, some decrease in the value of $\beta_{\rm d}$ can also be connected with the quantity $\Lambda = \Lambda_{0} -  3 \, {\rm ln}\Lambda_{0}$, where  $\Lambda_{0}$ is given by (\ref{48}). As one can see from Table~\ref{tab0}, for most pulsars located in 'the death valley', the values of $\Lambda$ are 35--40, while the normalization $\Lambda = 15$ in (\ref{betaMHD})--(\ref{betaBGI}) was given for ordinary pulsars ($P = 1$ s, ${\dot P}_{-15} = 2$). As a result, this reduction factor turns out to be  $\Lambda_{15}^{-1/2} \approx 0.6$.

As for the factor $h(x_{0})$, taking into account the distribution of the potential $\psi$ from the distance $x_{0}$ to the magnetic axis, it is easy to check that $h(x_{0}) \approx 3$ for \mbox{$0.5 < x_{0} < 0.6$.} Therefore, in what follows, we put
\begin{equation}
h(x_0) \approx 3.1.    
\end{equation}

Finally, note a completely different dependence of the functions $F(\chi)$ (\ref{FMHD})--(\ref{FBGI}) on the angle $\chi$ (they are normalized so that $F(0) = 1$). If in the MHD model, the function $F(\chi)$ increases with increasing the angle $\chi$ (and, therefore, large angles $\chi$ do not help us explain the small values of $\beta_{\rm b}$), in the BGI model, the function $F(\chi)$ decreases with increasing $\chi$ reaching a minimum at $\chi \sim 90^{\circ}$. THe corresponding values of $F^{\rm BGI}$ are given in Table~\ref{tab2} for $x_{0} = 0.7$. Unfortunately, the inaccuracy in determining the coefficient $k$ in (\ref{CalC}) gives a significant spread in the values of $F^{\rm BGI}$. Nevertheless, it can be stated with certainty that here, too, the reducing factor can reach the values $0.3$--$0.4$. However, in what follows, we put
\begin{equation}
F^{\rm BGI} = 0.7,
\label{FFBGI}
\end{equation}
because this value will better fit the entire angle range $\chi$. We emphasize once again that the minimum values $F$ for the BGI model are achieved at large inclination angles $\chi \sim 90^{\circ}$, while in the MHD model, the smallest values of $F$ occur at angles $\chi$ close to $0^{\circ}$.

\begin{table}
\caption{Minimum values of the factor $F^{\rm BGI}$ (\ref{FBGI}) for  \mbox{$x_{0} = 0.7$.} The values in the parentheses show the appropriate inclination \mbox{angles $\chi$.}} 
\begin{tabular}{ccccccc}
\hline
$P({\rm s})$  & 0.5 & 1 & 2 & 4 & 8 & 16    \\
\hline
$k = 0.2$  & 0.38& 0.36 & 0.33 & 0.31 &  0.29 & 0.27 \\
  & ($84^{\circ}$) & ($86^{\circ}$) & ($86^{\circ}$) &  ($87^{\circ}$) & ($87^{\circ}$) & ($87^{\circ}$)\\
$k = 0.5$  & 0.51 & 0.47 & 0.44 & 0.41 & 0.38 & 0.35 \\
  & ($80^{\circ}$) & ($82^{\circ}$) & ($83^{\circ}$) &  ($84^{\circ}$) & ($85^{\circ}$) & ($86^{\circ}$)\\
$k = 1$  & 0.62 & 0.57 & 0.53 & 0.49 &  0.45& 0.42 \\
  & ($75^{\circ}$) & ($80^{\circ}$) & ($81^{\circ}$) &  ($82^{\circ}$) & ($84^{\circ}$) & ($84^{\circ}$)\\
$k = 2$  & 0.75  & 0.69 & 0.64 & 0.59 &  0.54 & 0.52 \\
  & ($70^{\circ}$) & ($75^{\circ}$) & ($77^{\circ}$) &  ($80^{\circ}$) & ($81^{\circ}$) & ($82^{\circ}$) \\
$k = 4$  & 0.90 & 0.83 & 0.77 & 0.71 &  0.65 & 0.60 \\
 & ($60^{\circ}$) & ($65^{\circ}$) & ($70^{\circ}$) &  ($75^{\circ}$) & ($77^{\circ}$)& ($80^{\circ}$) \\
\hline
\end{tabular}
\label{tab2}
\end{table}

\begin{figure}
	\begin{minipage}{0.9\linewidth}
		\center{\includegraphics[width=1.0\linewidth]{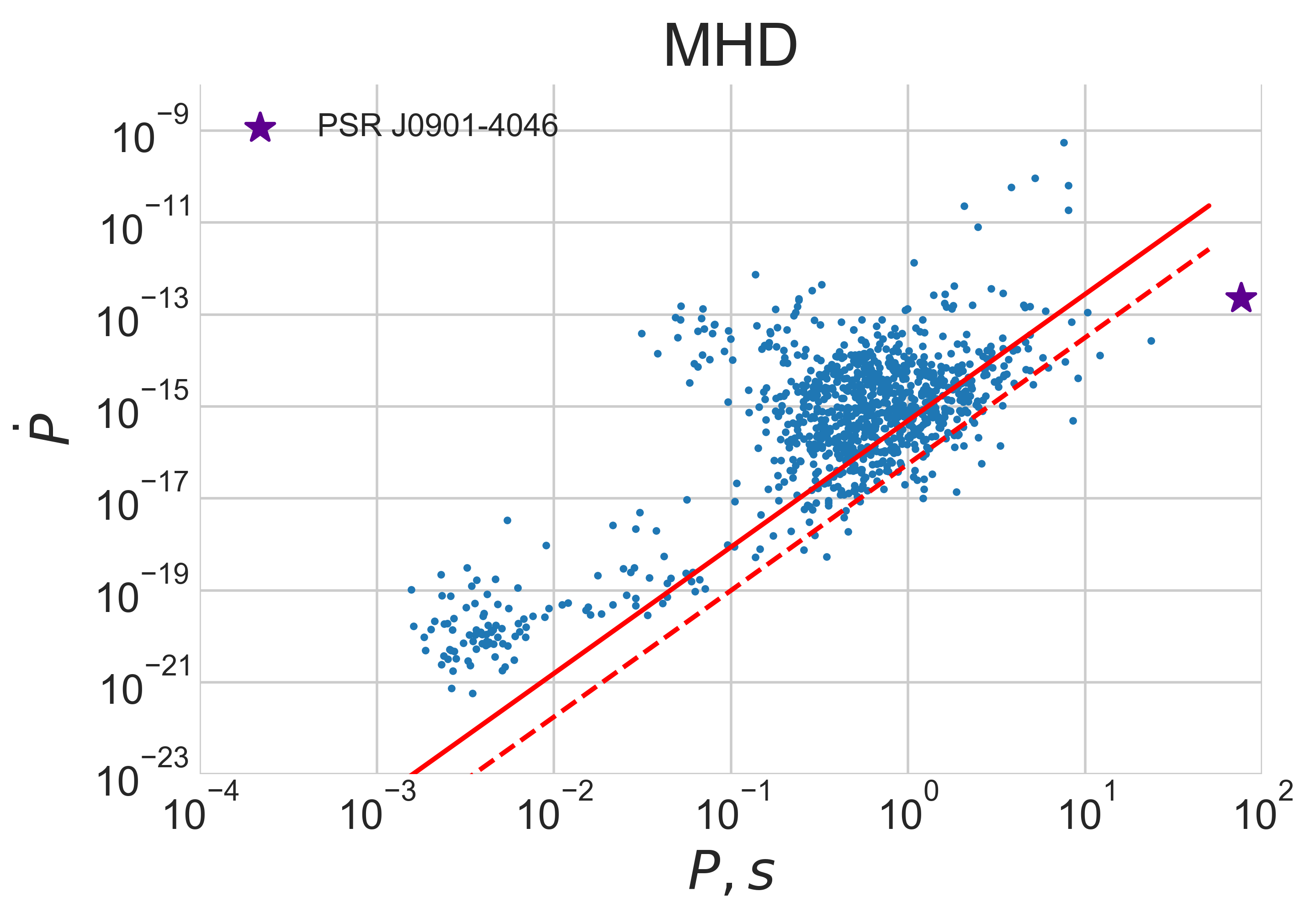}  }
	\end{minipage}
	\hfill
	\begin{minipage}{0.9\linewidth}
		\center{\includegraphics[width=1.0\linewidth]{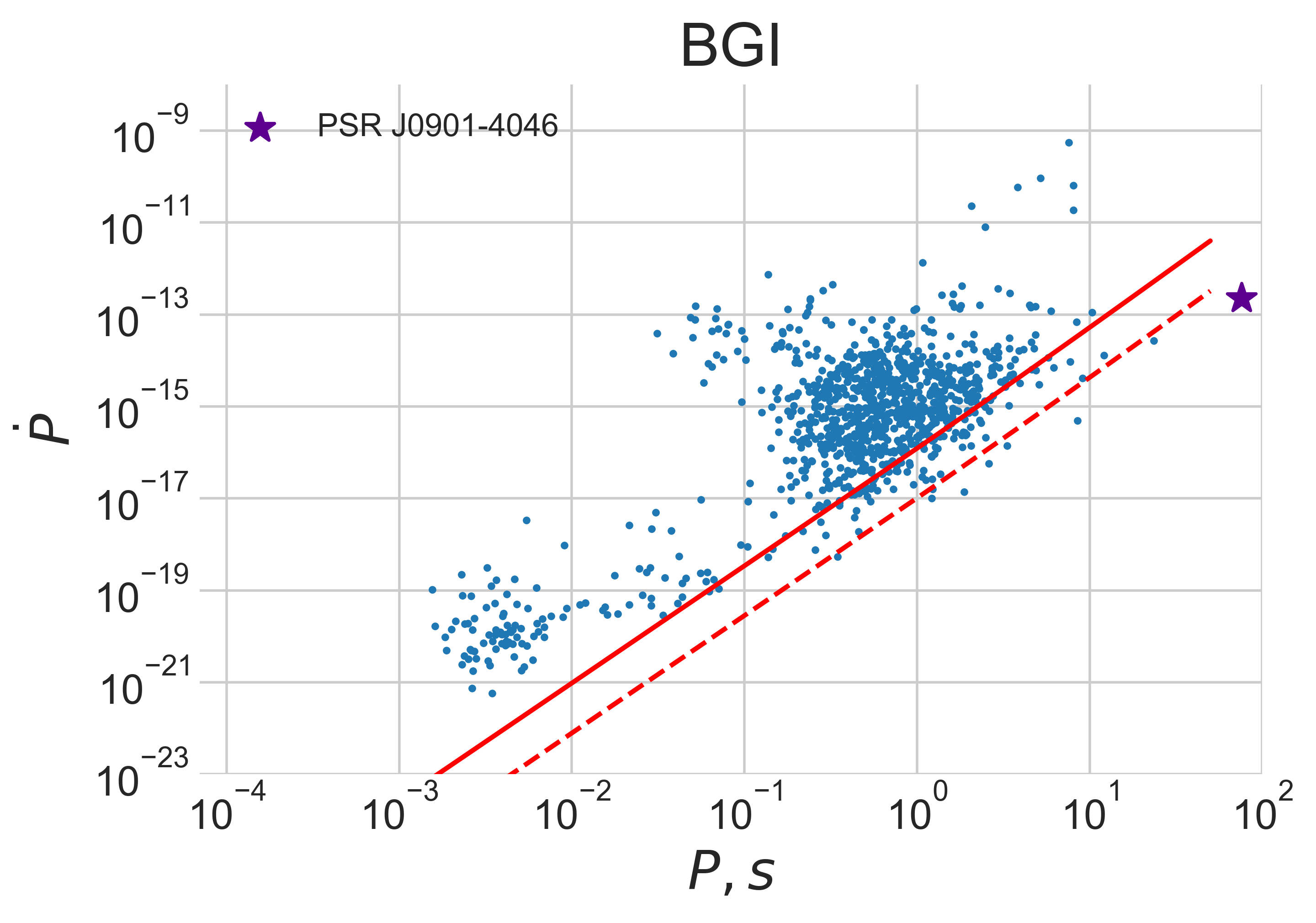}  }
	\end{minipage}
	\caption{'The death lines' for models MHD (top) and BGI (bottom). The solid lines correspond to the average value of the parameters in the expressions (\ref{betaMHD})--(\ref{betaBGI}), and the dashed lines correspond to their limiting values.} 
\label{fig2}	
\end{figure}

Figure~\ref{fig2} shows 'the death lines' for the models MHD (top) and BGI (bottom). The solid lines correspond to the average value of the parameters in expressions (\ref{betaMHD})--(\ref{betaBGI}) (\mbox{$\xi = 9$,} $\Lambda = 35$, $f_{\ast} = 1.6$, $K_{\rm g} = 0.35$, $F = 1$), and the dashed line corresponds to their limiting values ($\xi = 11, \Lambda = 41$, $f_{\ast} = 1.9$, $K_{\rm g} = 0.07$, $F^{\rm BGI} = 0.7$). A small break at small periods is associated with the dependence of $R_{0}$ on $P$. As we see, in general, both models quite well reproduce the lower boundary of 'the death valley'.

 Of course, long-period pulsars ($P > 3$ s) are of special interest, especially recently discovered pulsar J0901$-$4046 ($P \approx 46$ s,~\citealt{P76}). In particular, the question arises whether the slope of 'the death line' can be approximated by the dependence ${\dot P} = \beta_{\rm d}P^{11/4}$ considered here. In our opinion, the number of pulsars with periods $P > 3$ s located near the lower boundary of 'the death valley' is insufficient to speak of a change in its shape. On the other hand, it is useful to consider these pulsars in more detail.

\begin{table}
\caption{Slowly rotationg pulsars ($P > 3$ s) located deep below 'the death line' ($\beta_{\rm d} < 0.02$) taken from the ATNF catalogue~\citep{ATNF}. Theoretical values $\beta_{\rm d}$ correspond to the limiting parameters discussed above.}
\begin{tabular}{cccccc}
\hline
PSR & $P$  & ${\dot P}_{-15}$  & $\beta_{\rm d}$  & $\beta_{\rm d}^{({\rm MHD})}$ & $\beta_{\rm d}^{({\rm BGI})}$ \\
    & (s)   &  &  \\
\hline
J0250$+$5854    &  23.53 &  27.16  &  0.003 & 0.013 &  0.002 \\
J0418$+$5732    &  9.01  &  4.10   &  0.010 & 0.018 & 0.003 \\
J1210$-$6550    &  4.24  &  0.43   &  0.008 & 0.017 & 0.003\\
 J0901$-$4046   &  75.89  &  215.   &   0.001 &  0.013&  0.001\\
J1503$+$2111    &  3.32  &  0.14   &  0.005 & 0.016 & 0.003 \\
J2144$-$3933    &  8.51  &  0.50   &  0.001  & 0.014 & 0.002 \\
J2251$-$3711    &  12.12 &  13.10  &  0.014  & 0.016 & 0.003\\
\hline
\end{tabular}
\label{tab3}
\end{table}

Table~\ref{tab3} lists the data for six long-period pulsars. Theoretical values $\beta_{\rm d}^{({\rm MHD})}$ and $\beta_{\rm d}^{({\rm BGI})}$ for two models of evolution correspond to the limiting parameters discussed above. As we see, BGI model does not contradict the observational data (the limiting values of the parameters give even smaller values of $\beta_{\rm d}$ compared to the observed values). As for the  difference for MHD model, we discuss this issue in Section 5.

\section{'The death line' knee}

Before proceeding to the analysis of the obtained results, let us discuss  qualitatively one more property of 'the death line'. At the time of this writing, 3282 pulsars were already discovered~\citep{ATNF}. This rather rich statistics clearly shows that the line limiting from below the population of pulsars on the $P$--${\dot P}$-diagram has a break at \mbox{$P \approx 0.3$ s} (see Figure~\ref{fig1}). Here, we show that this break can be easily explained.

\begin{figure}
		\center{\includegraphics[width=0.9\linewidth]{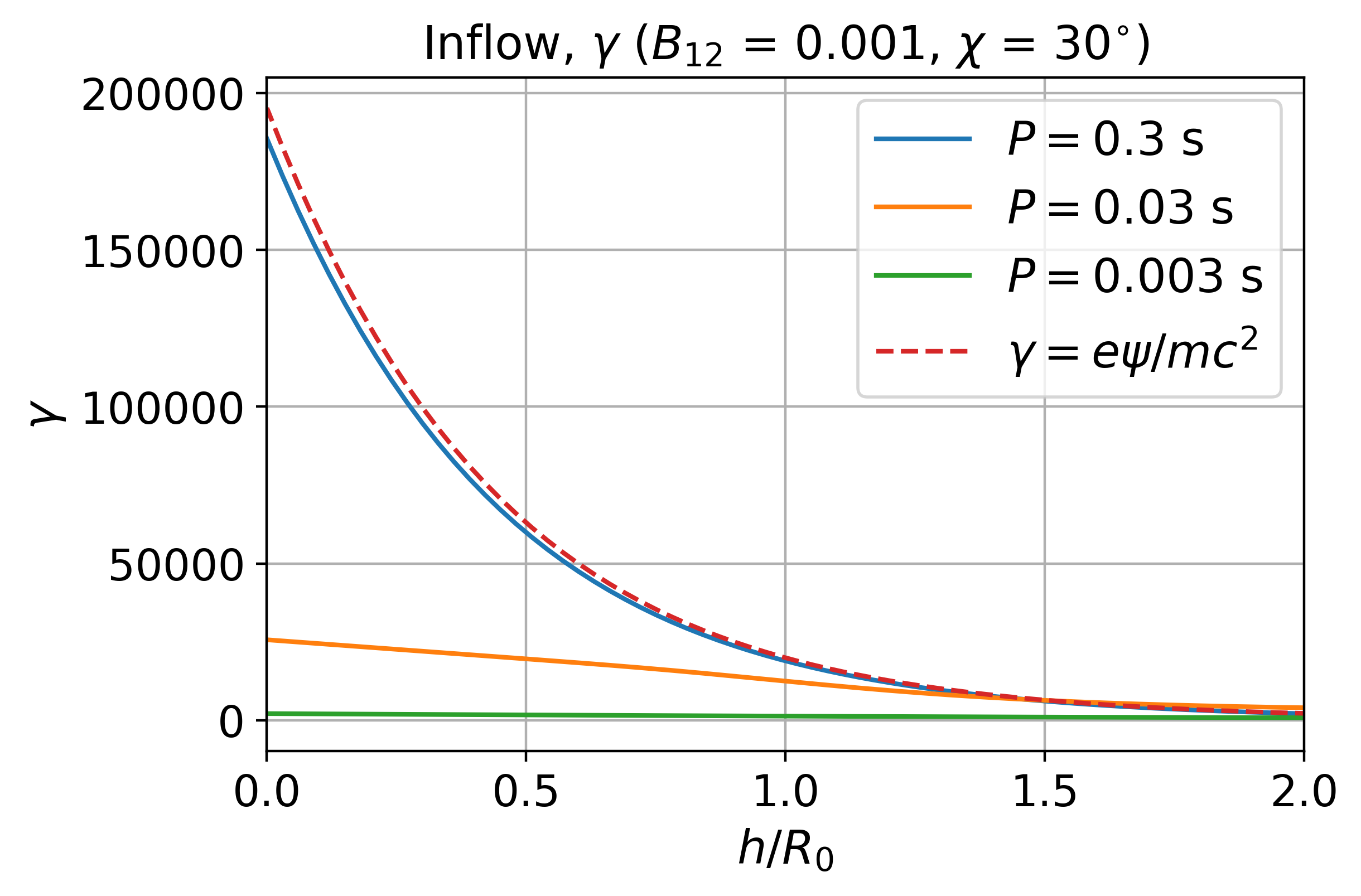}}
	\caption{Lorentz factor $\gamma(h)$ of back-moving primary particles depending on the distance $h$ from the star surface for three different periods $P = 0.003$ s, $P = 0.03$ s, and $P = 0.3$ s. The dashed line corresponds to the absence of the radiation reaction force associated with the curvature radiation.}
\label{fig3}	
\end{figure}

Indeed, as was shown in Paper I (see also~\citealt{Jones2022}), for the pulsars with  small enough periods (at any way, with periods $P < 0.1$ s), the radiation reaction becomes significant, so the energy of primary particles does not reach the values dictated by the potential drop $\psi$ (\ref{psis}). Clearly, this also applies to the back-moving primary particles. Figure~\ref{fig3} shows the dependence of the Lorentz-factors $\gamma(h)$ of the back-moving primary particles at the distance $h$ from the star surface for three different periods, $P = 0.003$ s, \mbox{$P = 0.03$ s,} and $P = 0.3$ s,  for the magnetic field $B = 10^{9}$ G which is characteristic of millisecond pulsars. The dashed line corresponds to the case when the radiation reaction force plays no role ($\gamma = e\psi/m_{\rm e}c^2$).

\begin{figure}
		\center{\includegraphics[width=0.9\linewidth]{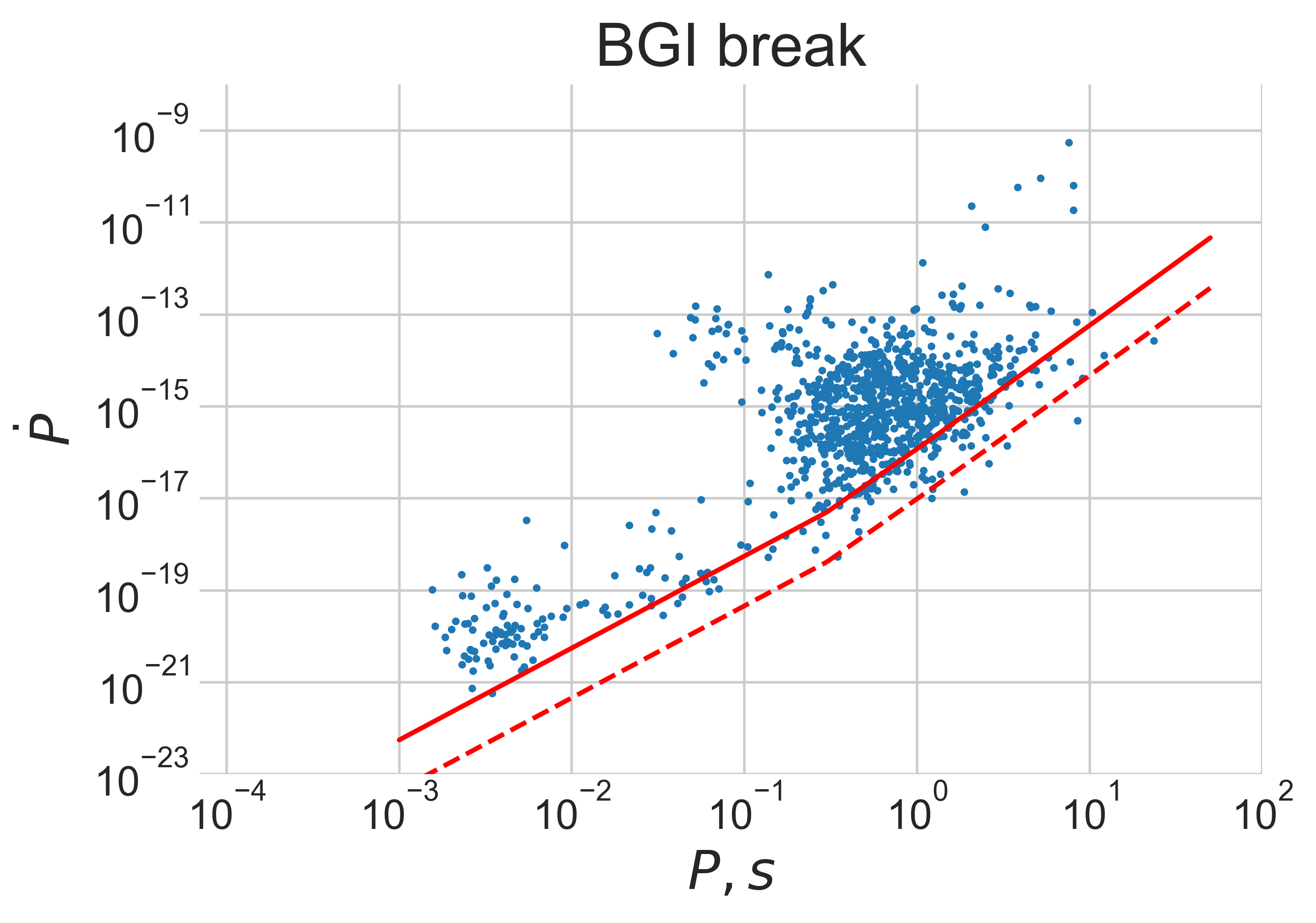}}
	\caption{'The death line' knee at period $P \approx 0.3$ s for the BGI model. At $P < 0.3$ s, the slope becomes noticeably flatter (it corresponds to proportionality ${\dot P} \propto P^2$).}
%	'Death line' obtained for the two models of pulsar braking. The dashed lines correspond to the 'classical' death line presented by~\citet{CR93}.}
\label{fig4}	
\end{figure}

As one can see, at $P < 0.3$ s, the energy of the primary particles becomes lower than previously assumed. Correspondingly, 'the death line' for these pulsars should be shifted upward compared to the dependence defined above. As a result, for the existence of cascade particle production, the corresponding rotation periods $P$ must be noticeably longer compared to the case in which the particle energy exactly corresponds to the accelerating potential $\psi$. And this, in turn, should lead to a rise in the death line in comparison with the asymptotic behavior corresponding to the periods $P > 0.3$ s.

 To evaluate this effect, one can use relation (\ref{Pmax}), in which the magnetic field $B$ should be considered as a function of $P$ and ${\dot P}$, and we also need to replace ${\cal P}$ by $k{\cal P}$ where the coefficient $k$ (defined for given magnetic field $B(P, {\dot P})$, as in Figure~\ref{fig3}) is the decrease in particle energy due to radiation reaction 
\begin{equation}
k = \frac{\gamma(0)}{\gamma_{\psi}(0)}.
\label{k}
\end{equation}
Here $\gamma_{\psi}(0)$ is the Lorentz-factor of the particles with the absence of the energy losses. The resulting relation implicitly determines the dependence ${\dot P} = {\dot P}(P)$ for 'the death line'.

%\begin{equation}
%P_{\rm max} = 0.7 \, \xi^{2/15} \frac{K_{\psi}^{2/5}}{K_{\rm cur}^{4/15}}
%f_{1.6}^{3/5}
%\Lambda_{15}^{2/15}
%R_{12}^{19/15}
%B_{12}^{8/15}x_{0}^{4/15} (k{\cal P})^{2/5} \, {\rm s}.
%\label{Pmaxknee}
%\end{equation}

The corresponding break of 'the death valley' for the model BGI is shown in Figure~\ref{fig4}. 
%the position of 'the death line' for the two models of pulsar braking. The dashed lines correspond to 'the classic death line'~\citep{CR93}. 
As one can see, at $P < 0.3$ s, the slope becomes noticeably flatter (it corresponds to proportionality ${\dot P} \propto P^2$). Herewith, such 'the death valley' corresponds even better to the observations. A more detailed discussion of this issue is beyond the scope of this work.

\section{Discussion and conclusion}

Thus, it was shown that 'the death valley' in the $P$--${\dot P}$ diagram is wide enough to explain all the observed sources even for a dipole magnetic field. In this case, the best agreement takes place in the BGI model. Indeed, for the limiting values of the parameters (($\xi = 11, \Lambda = 41$, $f_{\ast} = 1.9$, $K_{\rm g} = 0.07$, $F^{\rm BGI} = 0.7$), we get $\beta_{\rm b} = 0.003$, which allows us to explain almost all the sources collected in Table~\ref{tab0}. However, in our opinion, it is not worth arguing that the MHD model is inconsistent with the observational data. After all, the discrepancy here is only in factor 3 ($\beta_{\rm d} = 0.015$ for the above critical parameters, but with \mbox{$F^{\rm MHD} = 1$),} which can be associated with many reasons not taken into account in this work.

First of all, this difference can be related to a non-dipole magnetic field, which, as is well-known~\citep{Arons93, AKh2002, BTs2010, IEP2016}, leads to a decrease in the curvature radius of the magnetic field lines $R_{\rm c}$. As can be seen from relations (\ref{Kg}) and (\ref{betaMHD})--(\ref{betaBGI}), the corresponding factor $K_{\rm cur}$ enters linearly into the expression for $\beta_{\rm d}$. Hence, a decrease in the curvature radius $R_{\rm c}$ by only a few times makes it possible to explain many sources located in the lower part of 'the death valley'.

The second possibility is related to the size of the polar cap, the dependence on which is determined by the value $f_{\ast}$. A strong dependence on this parameter makes it possible to significantly reduce the value of $\beta_{\rm d}$ by a factor of three at a value of $f_{\ast} = 3$, which corresponds to an increase in the radius of the polar cap $R_{0}$ only by 20\% compared to the value $f_{\ast} = 1.9$ used above. Because we are unlikely to know the value of $f_{\ast}$ with such accuracy, increasing the value of this parameter can also lower the parameter $\beta_{\rm d}$ in the MHD model.

\begin{table}
\begin{center}
\caption{Intermittent pulsars}
\begin{tabular}{cccc}
\hline
PSR  & $P$ (s) & ${\dot P}_{-15}$ & ${\dot \Omega}_{\rm on}/{\dot \Omega}_{\rm off}$ \\
\hline
J1832+0029 & 0.53 & 1.55 &  1.5 \\
J1841+0500 & 0.91 & 34.7 & 2.5 \\
J2310+6706 & 0.81 & 8.11 &  1.8 \\
\hline
\end{tabular}
\end{center}
\label{tab4}
\end{table}

There may be other reasons leading to a decrease in the value of ${\dot P}$. In Table~\ref{tab4}, we collect three intermittent pulsars  for which the deceleration rates both in on and off regime are known (see~\citealt{BNhalf, GI} for more detail; more numerous pulsars with short nullings make it impossible to determine this ratio). As one can see, in the off state, the deceleration rate of the pulsar can be 1.5--2.5 times less than in the on state. Accordingly, the long-time averaged deceleration rate ${\dot P}$ may be less than we assume.

Summing up, it was shown that 'the death valley' in the $P$--${\dot P}$ diagram is wide enough to explain all the observed sources even for a dipole magnetic field. In this case, the best agreement takes place in the BGI model, although MHD model, taking into account quite reasonable additional assumptions, also does not contradict the observations. This once again proves that from the very beginning (i.e. from the works of~\citealt{sturrock, RS}) we correctly understood the nature of the activity of radio pulsars.

\section*{Data availability}
The data underlying this work will be shared on reasonable   request to the corresponding author.

\section*{Acknowledgements}
The authors thank Ya.N.Istomin and A.A.Philippov for their useful discussions. This work was partially 
supported by Russian Foundation for Basic Research (RFBR), grant 20-02-00469. 

\bibliographystyle{mnras}
\bibliography{references}

\end{document}